\documentclass[reprint,noshowpacs,noshowkeys,prd,balancelastpage,nofootinbib]{revtex4}
\usepackage[utf8]{inputenc}
\usepackage{color}
\usepackage{amsmath}
\usepackage{amssymb}
\usepackage{graphicx}
\usepackage[unicode=true,
 bookmarks=false,
 breaklinks=false,pdfborder={0 0 1},backref=section,colorlinks=true]
 {hyperref}
\hypersetup{
 linkcolor=purple,urlcolor=purple,citecolor=blue}

\makeatletter
\@ifundefined{textcolor}{}
{%
 \definecolor{BLACK}{gray}{0}
 \definecolor{WHITE}{gray}{1}
 \definecolor{RED}{rgb}{1,0,0}
 \definecolor{GREEN}{rgb}{0,1,0}
 \definecolor{BLUE}{rgb}{0,0,1}
 \definecolor{CYAN}{cmyk}{1,0,0,0}
 \definecolor{MAGENTA}{cmyk}{0,1,0,0}
 \definecolor{YELLOW}{cmyk}{0,0,1,0}
}

\pdfpageheight\paperheight
\pdfpagewidth\paperwidth

\newenvironment{lyxlist}[1]
	{\begin{list}{}
		{\settowidth{\labelwidth}{#1}
		 \setlength{\leftmargin}{\labelwidth}
		 \addtolength{\leftmargin}{\labelsep}
		 }}
	{\end{list}}

\pdfoutput=1

\usepackage{amsfonts}
\usepackage{mdframed}
\usepackage{footnote}
\usepackage{float}
\usepackage[font=footnotesize,it]{caption}
\usepackage{tikz}
\usepackage{tikz-3dplot}

\makeatother

\begin{document}
\title{Thin-shell Wormholes with Ordinary Matter in Pure Gauss-Bonnet Gravity}
\author{S. Danial Forghani}
\email{danial.forghani@final.edu.tr}

\affiliation{Faculty of Engineering, Final International University, Kyrenia, North
Cyprus via Mersin 10, Turkey}
\author{S. Habib Mazharimousavi}
\email{habib.mazhari@emu.edu.tr}

\affiliation{Department of Physics, Faculty of Arts and Sciences, Eastern Mediterranean
University, Famagusta, North Cyprus via Mersin 10, Turkey}
\begin{abstract}
In this paper, we introduce higher dimensional thin-shell wormholes
in pure Gauss-Bonnet gravity. The focus is on thin-shell wormholes
constructed by $N\geq5$-dimensional spherically symmetric vacuum
solutions. The results suggest that, under certain conditions, it
is possible to have thin-shell wormholes that both satisfy the weak
energy condition and be stable against radial perturbations. 
\end{abstract}
\date{\today }
\maketitle

\section{Introduction}

Wormholes and black holes are the most interesting solutions of the
Einstein's theory of gravity. While black holes are attractive for
their simple structure and existence of the so-called event horizon,
wormholes are rather mysterious for their geometrical and topological
structures \cite{WH}. Wormholes are hypothetical passages between
two distinct and distant points within the same or different spacetimes.
Traversable wormholes are even more interesting due to the opportunity
they provide a traveler to make, in principle, impossible journeys
possible \cite{WH}. Due to the structure of a wormhole solution in
$R$-gravity (general relativity), the corresponding energy-momentum tensor does not satisfy
the necessary energy conditions \cite{WH}, and hence, traversable
wormholes are considered as ``exotic'' spacetimes. The major energy
condition to determine the ordinariness of matter in wormhole literature
is the weak energy condition (WEC). The WEC states $T_{\mu\nu}V^{\mu}V^{\nu}\geq0$,
in which $T_{\mu\nu}$ is the energy-momentum tensor and $V^{\mu}$
is an arbitrary timelike vector. In the context of perfect fluids,
the WEC translates to two simultaneous conditions $\rho\geq0$ and
$\rho+p\geq0$, where $\rho$ is the energy density and $p$ is the
pressure\footnote{In the manuscript, since we deal with the surface energy density of
thin-shell wormhole, we symbolize the energy density by $\sigma$
instead of $\rho$.}. The latter condition alone is the null energy condition (NEC), which
is clearly implied by the WEC. Since all known matters satisfy the
WEC, we have given the name ``exotic'' to matters who do not. Believing
in the nonexistence of exotic matters in our universe implies that
traversable wormholes are not physical but some mathematical entities.
Any attempt for finding wormhole solutions supported by ordinary matter
in $R$-gravity has failed from the beginning. Hence, researchers
moved on with modified theories of gravity such as $f\left(R\right)$
and Lovelock theories \cite{LWH}.

On the other hand, an attempt by Visser to construct traversable wormholes
using junction formalism brought some hope to the wormhole community
since it minimized the amount of the exotic matter (in case they are
not avoidable) \cite{TSW}. Such wormholes have been called thin-shell
wormholes (TSWs). Initially, TSW was proposed in Einstein's $R$-gravity
where two identical flat spacetimes with holes were glued together
at the boundary of their holes. The common hole was indeed the throat
of the constructed TSW and provided a passage between the two flat
spacetimes \cite{TSW}. Having finely chosen the geometry of the throat,
gives the possibility to minimize the exotic matter which presents
at the throat. The concept has been developed over the last three
decades such that a rich literature on different aspects of TSWs are
available \cite{TSWLit}. Although TSWs are different from the former
classic wormholes - due to the fact that they are not direct solutions
to the Einstein field equation - they also suffer from the same obstacle
as their former. This means that TSWs in $R$-gravity are supported
by exotic matters, no matter what.

Moreover, TSWs may also suffer from instability against an external
perturbation \cite{Stab}. This is an important issue due to the application
of TSWs. A traveler (or signal) who uses the throat, in general, makes
interaction with the TSW which can be considered as a small or large
perturbation. If such wormhole is not stable against the perturbation,
it either collapses or evaporates. This is why, almost all constructed
TSWs in the literature have been investigated for their stability,
as well. 

Here the question is ``will constructing TSWs in modified theories
of gravity give chances for having TSWs supported by ordinary matter?''
The answer is yes \cite{TSWL1,TSWL2,TSWL3,TSWL4,TSWL5,TSWL6} and
in the present study we shall give another evidence for a positive
answer. In particular, in \cite{TSWL1}, authors study a TSW and its
stability in $5$-dimensional Gauss--Bonnet (GB) gravity augmented
by a Maxwell electromagnetic field, and find that the additional GB
Lagrangian not only broadens the range of possible stable regions
but also limits the amount of required exotic matter at the throat.
Using the same gravity, authors show in \cite{TSWL2} that with fine-tuning
the parameters, one could construct a TSW supported by ordinary matter,
only when the GB parameter is negative. In \cite{TSWL3}, the radial
stability of this TSW is investigated and it is shown that it is stable
only for a very narrow region of fine-tuned parameters, again, only
in case of a negative GB parameter. Another higher dimensional stable
TSW with ordinary matter is studied in Einstein--Yang--Mills--Gauss--Bonnet
gravity in \cite{TSWL4}. This TSW also exhibits the same behavior,
in the sense that it is supported by ordinary matter only for negative
values of GB parameter, and is stable against radial perturbation
only for a narrow region on the stability diagram. In another attempt,
authors construct a TSW in third-order Lovelock gravity in \cite{TSWL5},
and indicate that for a negative GB parameter and a positive third-order
Lovelock parameter, the TSW could be sustained by ordinary matter.
Yet, the stability diagrams show that for such TSW, the throat is radially
stable only when the sound speed is negative throughout the matter.
Finally, in \cite{TSWL6}, authors study the effect of the GB parameter
in the stability of TSWs supported by a generalized Chaplygin gas
and a general barotropic fluid, and conclude that although the GB
parameter highly affects the stability regions in the diagrams, yet
the corresponding TSWs does not satisfy the WEC. What distinguishes
the present paper from its ancestors can be listed as follows: i)
While in all previous studies an Einstein Lagrangian exists in the
action, here we study pure GB which is free of the Einstein term.
ii) In the mentioned studies above, a TSW could be supported by ordinary
matter only when the GB parameter was negative. However, a negative
GB parameter admits an exotic bulk spacetime where the associated
solutions do not imply the classic limits of Schwarzschild or Reissner--Nordström
solutions. On the other hand, in the present study the GB parameter
has to be positive due to the nature of the solutions considered.
iii) Unlike the previous studies, we do not limit our results by fine-tuning
the parameters. The solution we are about to employ is chargeless,
we consider no cosmological constant, and the GB parameter is positive-definite.
Yet, we find TSWs with ordinary matter for a wide range of TSW radius.
iv) We do not limit ourselves to a particular equation of state (EoS).
The two EoSs we have used here are the barotropic EoS and the variable
EoS \cite{Varela}. While the former is the most used EoS in TSW literature,
the latter is the most general EoS.

This terminology ``pure Lovelock gravity'' was used for the first
time by Kastor and Mann \cite{Kastor1} and has been developed by
Cai, \textit{et al.} \cite{Cai1}, Dadhich, \textit{et al.} \cite{PLT}
and others \cite{Others}. The interesting fact about pure Lovelock
gravity is that it admits non-degenerate vacua in even dimensions
and unique non-degenerate dS and AdS vacua in odd dimensions \cite{Cai1}.
Moreover, the corresponding black hole solutions are asymptotically
indistinguishable from the ones in Einstein gravity \cite{PLT}. This
similar asymptotic behavior of two theories seems to extend also to
the level of the dynamics and a number of physical degrees of freedom
in the bulk \cite{PLT}. The pure GB gravity that we consider here,
is the second--order pure Lovelock gravity.

The paper is arranged as follows. In section \ref{sec II} we briefly
review the Lovelock and the pure Lovelock gravity and their solutions.
In section \ref{sec III}, within the standard framework of thin-shell
formalism, we construct the TSW in the pure GB gravity for $N\geq5$
dimensions, and investigate the conditions under which the TSW could
be supported by ordinary matter. Section \ref{sec IV} is devoted
to the stability of the TSW against radial perturbation to see whether
our ordinary-mattered TSW could be stable or not. Unfortunately, to
the best of our knowledge, no method or formalism has been developed
so far to study the stability of a TSW against an angular perturbation.
Therefore, we only settle for a radial perturbation in this section.
Finally, we bring our conclusion in section \ref{sec V}. Throughout
the paper, we have used the convention $G_{N}=c=1$.

\section{Pure Lovelock gravity: a review \label{sec II}}

Lovelock theory, is one of the higher dimensional modified theories
of gravity which leaves the gravitational field equations second order
\cite{Lovelock}. The first order Lovelock theory is the Einstein
$R$-gravity in all dimensions. The second order Lovelock theory is
known as the Gauss-Bonnet (GB) theory and is defined for spacetimes
with dimensions of five and higher. The third order Lovelock theory
is applicable to seven dimensions and higher, and is well-known for
the two additional coupling constants it provides. The general vacuum
$N$-dimensional Lovelock theory is formulated with the action
\begin{equation}
I=\frac{1}{2\kappa_{N}}\int d^{N}x\sqrt{-g}\sum_{k=0}^{\left[\frac{N-1}{2}\right]}c_{k}\mathcal{L}_{k}\label{action}
\end{equation}
in which $\kappa_{N}$ is the Einstein's constant in $N$ dimensions,
$c_{k}$ are arbitrary real constants, $\left[\frac{N-1}{2}\right]$
is the integral part of $\frac{N-1}{2}$ and 
\begin{equation}
\mathcal{L}_{k}=\frac{1}{2^{k}}\delta_{\mu_{1}\nu_{1}...\mu_{k}\nu_{k}}^{\alpha_{1}\beta_{1}...\alpha_{k}\beta_{k}}\prod\limits _{i=1}^{k}R_{\text{ \ \ \ \ \ \ \ }\alpha_{i}\beta_{i}}^{\mu_{i}\nu_{i}}\label{Euler density}
\end{equation}
are the Euler densities of a $2k$-dimensional manifold, where the
generalized Kronecker delta $\delta$ is defined as the anti-symmetric
product 
\begin{equation}
\delta_{\mu_{1}\nu_{1}...\mu_{k}\nu_{k}}^{\alpha_{1}\beta_{1}...\alpha_{k}\beta_{k}}=k!\delta_{[\mu_{1}}^{\alpha_{1}}\delta_{\nu_{1}}^{\beta_{1}}...\delta_{\mu_{k}}^{\alpha_{k}}\delta_{\nu_{k}]}^{\beta_{k}}.\label{generalized Kronecker delta}
\end{equation}
For $k=0,$ we get $\mathcal{L}_{0}=1$ and $c_{0}$ will be the bare
cosmological constant. For $k=1,$ one finds the Einstein-Hilbert
Lagrangian where $\mathcal{L}_{1}=R$ and $c_{1}=1.$ The well known
GB Lagrangian is found with $k=2$ such that 
\begin{equation}
\mathcal{L}_{2}=\mathcal{L}_{GB}=R^{\kappa\lambda\mu\nu}R_{\kappa\lambda\mu\nu}-4R^{\mu\nu}R_{\mu\nu}+R^{2},\label{GB Lagrangian}
\end{equation}
where $c_{2}$ is called the GB parameter. Finally, the third order
Lovelock Lagrangian is given with $k=3$ where 
\begin{multline}
\mathcal{L}_{3}=2R^{\kappa\lambda\rho\sigma}R_{\rho\sigma\mu\nu}R_{\hspace{0.3cm}\kappa\lambda}^{\mu\nu}+8R_{\hspace{0.3cm}\kappa\lambda}^{\mu\nu}R_{\hspace{0.3cm}\nu\rho}^{\kappa\sigma}R_{\hspace{0.3cm}\mu\sigma}^{\lambda\rho}+24R^{\kappa\lambda\mu\nu}R_{\mu\nu\lambda\rho}R_{\hspace{0.15cm}\kappa}^{\rho}+3RR^{\kappa\lambda\mu\nu}R_{\kappa\lambda\mu\nu}\\
+24R^{\kappa\lambda\mu\nu}R_{\mu\kappa}R_{\nu\lambda}+16R^{\mu\nu}R_{\nu\sigma}R_{\hspace{0.15cm}\mu}^{\sigma}-12RR^{\mu\nu}R_{\mu\nu}+R^{3},\label{third-order Lagrangian}
\end{multline}
and $c_{3}$ is the third order Lovelock parameter.

Considering an $N$-dimensional spherically symmetric static spacetime
with line element 
\begin{equation}
ds^{2}=-A\left(r\right)dt^{2}+\frac{dr^{2}}{A\left(r\right)}+r^{2}d\Omega_{N-2}^{2},\label{metric}
\end{equation}
the Einstein-Lovelock's field equation reduces to a $k$-order ordinary
equation given by 
\begin{equation}
\sum_{k=0}^{\left[\frac{N-1}{2}\right]}\tilde{c}_{k}\psi^{k}=\frac{\mu}{r^{N-1}},\label{ordinary equation}
\end{equation}
in which $A\left(r\right)=1-r^{2}\psi\left(r\right)$, and the dimension-dependent
mass parameter $\mu$ is related to the ADM mass $M$ of the (possible)
asymptotically flat black hole or non-black hole solution by 
\begin{equation}
\mu=\frac{2\kappa_{N}M}{\left(N-2\right)\Sigma_{N-2}}.\label{mass parameter}
\end{equation}
Furthermore, 
\begin{equation}
\Sigma_{N-2}=\frac{2\pi^{\frac{N-1}{2}}}{\Gamma\left(\frac{N-1}{2}\right)}\label{surface area of a unit sphere}
\end{equation}
is the surface area of the $\left(N-2\right)$-dimensional unit sphere,
$\tilde{c}_{0}=\frac{c_{0}}{\left(N-1\right)\left(N-2\right)}$, $\tilde{c}_{1}=1$
and for $k\geq2$ 
\begin{equation}
\tilde{c}_{k}=\prod\limits _{i=3}^{2k}\left(N-i\right)c_{k}.\label{c-tilde parameters}
\end{equation}
In contrast to the general $m$-order Lovelock gravity with $1\leq m\leq\left[\frac{N-1}{2}\right]$,
in $m$-order pure Lovelock gravity \cite{Kastor1,Cai1,PLT,Others},
except for $c_{0}$, all $c_{k}$ for $k\neq m$ are zero, while $c_{k=m}\neq0.$
With the same line element as (\ref{metric}), the field equation
of the $m$-order pure Lovelock gravity becomes 
\begin{equation}
\tilde{c}_{0}+\tilde{c}_{m}\psi^{m}=\frac{\mu}{r^{N-1}}\label{the field equation of the m-order pure Lovelock gravity}
\end{equation}
where the general solution for $\psi$ is obtained to be 
\begin{equation}
\psi\left(r\right)=\left\{ \begin{array}{cc}
\pm\left[\frac{1}{\tilde{c}_{m}}\left(\frac{1}{\ell^{2}}+\frac{\mu}{r^{N-1}}\right)\right]^{\frac{1}{m}}, & m\text{ even}\\
\left[\frac{1}{\tilde{c}_{m}}\left(\frac{1}{\ell^{2}}+\frac{\mu}{r^{N-1}}\right)\right]^{\frac{1}{m}}, & m\text{ odd}
\end{array}\right.,\label{general solution of psi}
\end{equation}
in which $\ell$ is the cosmological length in $\tilde{c}_{0}=-\frac{1}{\ell^{2}}.$
Finally, the metric function is given by 
\begin{equation}
A\left(r\right)=\left\{ \begin{array}{cc}
1\mp r^{2}\left[\frac{1}{\tilde{c}_{m}}\left(\frac{1}{\ell^{2}}+\frac{\mu}{r^{N-1}}\right)\right]^{\frac{1}{m}}, & m\text{ even}\\
1-r^{2}\left[\frac{1}{\tilde{c}_{m}}\left(\frac{1}{\ell^{2}}+\frac{\mu}{r^{N-1}}\right)\right]^{\frac{1}{m}}, & m\text{ odd}
\end{array}\right..\label{general solution of the metric function A}
\end{equation}
In the rest of the paper we consider the pure GB gravity without the
cosmological constant by setting $m=2$ and $c_{0}=0$ which result
in 
\begin{equation}
A\left(r\right)=1\mp\omega^{2}\,r^{\left(5-N\right)/2},\label{metric function for GB}
\end{equation}
where $\omega^{2}\equiv\sqrt{\frac{\mu}{\tilde{c}_{2}}}$ is a positive
constant. This solution for $N=5$ and $N>5$ admits different asymptotic
behaviors. For $N=5$ the metric function in (\ref{metric function for GB})
reduces to 
\begin{equation}
A\left(r\right)=1\mp\omega^{2},\label{metric function for GB and N=00003D00003D5}
\end{equation}
constraint by $1\mp\sqrt{\frac{\mu}{\tilde{c}_{2}}}>0,$\textcolor{red}{{}
}which is an asymptotically non-flat spacetime and possesses a singularity
at $r=0$ with a conical structure accompanied by a deficit (surplus)
angle for the minus (plus) sign. For $N>5$ the asymptotically flat
solution in (\ref{metric function for GB}) has a singularity at $r=0$,
which is naked for the plus sign and is hidden behind an event horizon
located at 
\begin{equation}
r_{+}=\omega^{4/\left(N-5\right)}\label{event horizon}
\end{equation}
for the minus sign.

\section{Thin-shell wormholes in pure GB gravity \label{sec III}}

To construct a TSW in an $N$-dimensional $m$-order pure Lovelock
gravity, we excise out the inner part of a timelike hypersurface $\varSigma:=r-a=0$
in which $a>r_{h}$ ($r_{h}$ is the possible event horizon) and make
two identical copies from the rest of the bulk spacetime (\ref{metric}),
namely $\mathcal{M}^{\left(\pm\right)}.$ Afterwards, we glue the
two incomplete manifolds $\mathcal{M}^{\left(\pm\right)}$ at their
common boundary hypersurface $\varSigma.$ The resultant manifold,
i.e. $\mathcal{M}=\mathcal{M}^{\left(+\right)}\cup\mathcal{M}^{\left(-\right)}$,
is geodesically complete with a throat located at $r=a.$ Joining
the two incomplete manifolds at $\varSigma$ requires the so-called
generalized junction conditions to be satisfied. These conditions
are, in summary, as follows. First of all, the induced metric tensor
of the throat should be continuous across the shell i.e., 
\begin{equation}
\left[h_{ab}\right]_{-}^{+}=0\label{first junction condition}
\end{equation}
in which $\left[X\right]_{-}^{+}=\left(X\right)_{+}-\left(X\right)_{-}$
and $\left(h_{ab}\right)_{\pm}$ are the induced metric tensor at
either sides of the throat defined by 
\begin{equation}
\left(h_{ab}\right)_{\pm}\equiv\left(g_{\alpha\beta}\frac{\partial x^{\alpha}}{\partial\xi^{a}}\frac{\partial x^{\beta}}{\partial\xi^{b}}\right)_{\pm}.\label{induced metric}
\end{equation}
Herein, $\left(x^{\alpha}\right)_{\pm}=\left\{ t,r,\theta_{1},...,\theta_{N-2}\right\} _{\pm}$
are the coordinates of the bulk spacetime while $\left(\xi^{a}\right)_{\pm}=\left\{ \tau,\theta_{1},...,\theta_{N-2}\right\} _{\pm}$
are the coordinates of the hypersurface with $\tau$ being the proper
time. Upon satisfying the first junction condition, one finds $r_{\pm}=a\left(\tau\right),$
$\theta_{a\pm}=\theta_{a}$ and 
\begin{equation}
\dot{t}_{\pm}^{2}=\frac{1}{A\left(a\right)}\left(1+\frac{\dot{a}^{2}}{A\left(a\right)}\right),\label{result of the first junction condition}
\end{equation}
in which a dot stands for a derivative with respect to the proper
time $\tau.$ Hence, the induced metric of the throat becomes 
\begin{equation}
ds_{\varSigma}^{2}=-d\tau^{2}+a^{2}d\Omega_{N-2}^{2}.\label{metric of the shell}
\end{equation}
The second junction condition implies that there is a discontinuity
at the throat associated with the energy-momentum tensor of the fluid
at the throat, given by the equation \cite{Davis} 
\begin{equation}
-\kappa_{N}S_{ab}=2c_{2}\left(3\left[J_{ab}\right]_{-}^{+}-h_{ab}\left[J\right]_{-}^{+}+2\left[\hat{P}_{amnb}K^{mn}\right]_{-}^{+}\right).\label{second junction condition}
\end{equation}
In (\ref{second junction condition}), $S_{a}^{b}=diag\left[-\sigma,p,p,...,p\right]$
is the surface energy-momentum tensor, 
\begin{equation}
\hat{P}_{amnb}=\hat{R}_{amnb}+\left(\hat{R}_{mn}h_{ab}-h_{an}\hat{R}_{mb}\right)-\left(\hat{R}_{an}h_{mb}-\hat{R}_{ab}h_{mn}\right)+\frac{1}{2}\hat{R}\left(h_{an}h_{mb}-h_{ab}h_{mn}\right)\label{tensor P}
\end{equation}
is the divergence-free part of the Riemann tensor $\hat{R}_{amnb}$
(compatible with the metric of the induced metric), 
\begin{equation}
J_{ab}=\frac{1}{3}\left(2KK_{am}K_{b}^{m}+K_{mn}K^{mn}K_{ab}-2K_{am}K^{mn}K_{nb}-K^{2}K_{ab}\right),\label{tensor J}
\end{equation}
and $J=J_{a}^{a}$. Furthermore, in (\ref{tensor J}) $K_{ab}$ is
the extrinsic curvature tensor (the second fundamental form) of the
hypersurface defined by 
\begin{equation}
\left(K_{ab}\right)_{\pm}=-\left(n_{\gamma}\right)_{\pm}\left(\frac{\partial^{2}x^{\gamma}}{\partial\xi^{a}\partial\xi^{b}}+\Gamma_{\alpha\beta}^{\gamma}\frac{\partial x^{\alpha}}{\partial\xi^{a}}\frac{\partial x^{\beta}}{\partial\xi^{b}}\right)_{\pm}\label{curvature tensor}
\end{equation}
with the spacelike normal vector given by 
\begin{equation}
\left(n_{\gamma}\right)_{\pm}=\pm\left(\frac{1}{\sqrt{g^{\alpha\beta}\frac{\partial\varSigma}{\partial x^{\alpha}}\frac{\partial\varSigma}{\partial x^{\beta}}}}\frac{\partial\varSigma}{\partial x^{\gamma}}\right)_{\pm}.\label{normal vector}
\end{equation}
Using (\ref{second junction condition}), one obtains the surface
energy density and the lateral pressures as \cite{Mazhari} 
\begin{equation}
\sigma=-\frac{4\left(N-2\right)\tilde{c}_{2}}{3\kappa_{N}a^{3}}\sqrt{A+\dot{a}^{2}}\left(3-A+2\dot{a}^{2}\right)\label{energy density}
\end{equation}
and 
\begin{equation}
p=\frac{4\tilde{c}_{2}}{3\kappa_{N}a^{2}\sqrt{A+\dot{a}^{2}}}\left[3\ddot{a}\left(1+A+2\dot{a}^{2}\right)+\frac{3}{2}A^{\prime}\left(1-A\right)+\frac{\left(N-5\right)}{a}\left(A+\dot{a}^{2}\right)\left(3-A+2\dot{a}^{2}\right)\right],\label{pressure}
\end{equation}
respectively. Note that, a dot stands for a derivative with respect
to $\tau$ while a prime implies a derivative with respect to the
radius $a$. By assuming a static equilibrium configuration for the
throat, where $a=a_{0}$ and $\dot{a}=\ddot{a}=0$, the static surface
energy density $\sigma_{0}$ and pressure $p_{0}$ are obtained as
\begin{equation}
\sigma_{0}=-\frac{4\left(N-2\right)\tilde{c}_{2}}{3\kappa_{N}a_{0}^{3}}\sqrt{A_{0}}\left(3-A_{0}\right)\label{static energy density}
\end{equation}
and 
\begin{equation}
p_{0}=\frac{4\tilde{c}_{2}}{3\kappa_{N}a_{0}^{2}\sqrt{A_{0}}}\left(\frac{3}{2}A_{0}^{\prime}\left(1-A_{0}\right)+\frac{\left(N-5\right)}{a_{0}}A_{0}\left(3-A_{0}\right)\right),\label{static pressure}
\end{equation}
respectively, where $A_{0}=\left.A\left(r\right)\right|_{r=a_{0}}$
and $A_{0}^{\prime}=\left.dA\left(r\right)/dr\right|_{r=a_{0}}$.
In what follows we shall study some specific cases regarding the dimensions
and black hole/non-black hole spacetimes.

\subsection{$N>5$-dimensional asymptotically flat bulk spacetime}

As we have mentioned previously, for $N>5$ in the asymptotically
flat solution (\ref{metric function for GB}), while the plus sign
represents a non-black hole solution the minus sign admits a black
hole. Inserting the static version of (\ref{metric function for GB})
into (\ref{static energy density}) and (\ref{static pressure}) one
obtains 
\begin{equation}
\sigma_{0}=-\frac{4\left(N-2\right)\tilde{c}_{2}}{3\kappa_{N}a_{0}^{3}}\sqrt{1\mp\omega^{2}\,a_{0}^{\left(5-N\right)/2}}\left(2\pm\omega^{2}\,a_{0}^{\left(5-N\right)/2}\right)\label{energy density for non-BL}
\end{equation}
and 
\begin{equation}
\sigma_{0}+p_{0}=\frac{4\tilde{c}_{2}}{\kappa_{N}a_{0}^{3}\sqrt{1\mp\omega^{2}\,a_{0}^{\left(5-N\right)/2}}}\left[\frac{N-1}{4}\left(\omega^{2}\,a_{0}^{\left(5-N\right)/2}\right)^{2}-2\pm\omega^{2}\,a_{0}^{\left(5-N\right)/2}\right],\label{pressure for non-BL}
\end{equation}
where the upper (lower) sign is corresponding to the black hole (non-black
hole) solution. For the lower sign, noting that $N>5$, we may have
$\sigma_{0}\geq0$ and $\sigma_{0}+p_{0}\geq0$ only if 
\begin{equation}
a_{0}\leq\left(\frac{\omega^{2}}{2}\right)^{2/\left(N-5\right)}=a_{c}.\label{critical radius}
\end{equation}
Therefore, for any throat radius equal to or smaller than this critical
radius $a_{c}$, a TSW constructed by an $N>5$-dimensional non-black
hole spacetime solution to the pure GB gravity satisfies the WEC;
consequently, the throat is indeed supported by ordinary matter rather
than exotic. On the other hand, it is evident from (\ref{energy density for non-BL})
that the energy density for the black hole solution is negative-definite.
Hence, a TSW constructed by such a spacetime is absolutely sustained
by exotic matter.

\subsection{$5$-dimensional conical bulk spacetime}

In $5$-dimensional pure GB gravity, $A\left(a_{0}\right)=A_{0}\equiv1\pm\omega^{2}$
is a positive constant, given by (\ref{metric function for GB and N=00003D00003D5}).
The solution is singular at $r=0$ and admits deficit/surplus angle
depending on whether $A_{0}$ is less or greater than unity, indicating
the existence of a cosmic string. Following (\ref{static energy density})
and (\ref{static pressure}), one finds the surface energy density
and lateral pressure at the throat by 
\begin{equation}
\sigma_{0}=-\frac{4\tilde{c}_{2}}{\kappa_{N}a_{0}^{3}}\sqrt{A_{0}}\left(3-A_{0}\right)\label{energy density for CS}
\end{equation}
and 
\begin{equation}
p_{0}=0.\label{Pressure for CS}
\end{equation}
Since $A_{0}$ is positive, for any $A_{0}\geq3$ we find $\sigma_{0}\geq0$
and $\sigma_{0}+p_{0}\geq0$. The matter at the throat satisfies the
WEC and so is ordinary.

\section{Stability analysis \label{sec IV}}

To study the stability of the TSW in pure GB gravity, we start with
the expression of $\sigma$ and $p$ given by Eqs. (\ref{energy density})
and (\ref{pressure}). From the energy conservation equation, i.e.
$S_{;b}^{ab}=0$, one finds that the energy density in (\ref{energy density})
and the pressure in (\ref{pressure}) satisfy the relation 
\begin{equation}
\frac{d\sigma}{da}+\frac{N-2}{a}\left(\sigma+p\right)=0.\label{energy conservation}
\end{equation}
In addition, Eq. (\ref{energy density}) can be written in the form
of a one-dimensional equation of motion for the radius of the throat
as 
\begin{equation}
\dot{a}^{2}+V\left(a\right)=0,\label{equation of motion}
\end{equation}
in which the first term is kinetic and the the second term is given
by the effective potential 
\begin{equation}
V\left(a\right)=A\left(a\right)-\left(\Psi\left(a\right)-\frac{1-A\left(a\right)}{2\Psi\left(a\right)}\right)^{2},\label{effective potential}
\end{equation}
where 
\begin{equation}
\Psi\left(a\right)=\left[\frac{\frac{16\left(N-2\right)\tilde{c}_{2}}{3\kappa_{N}a^{3}}}{\sqrt{\sigma^{2}+2\left(\frac{4\left(N-2\right)\tilde{c}_{2}}{3\kappa_{N}a^{3}}\right)^{2}\left(1-A\left(a\right)\right)^{3}}-\sigma}\right]^{1/3}.\label{psi}
\end{equation}

\begin{figure}[h]
\includegraphics[scale=0.3]{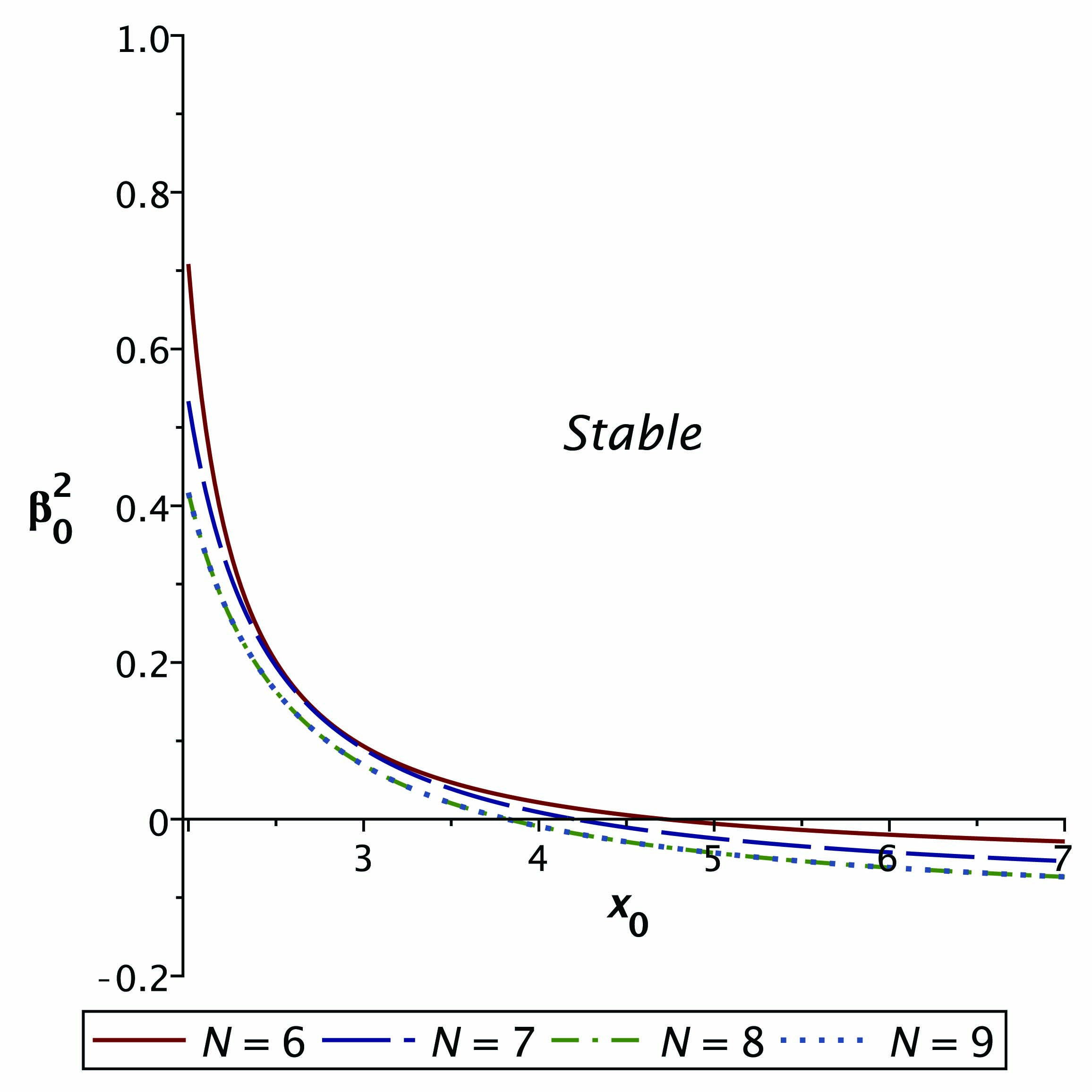}

\caption{The stability diagrams for the non-black hole solution and a barotropic
equation of state. The diagram plots $\beta_{0}^{2}$ against $x_{0}\equiv\omega^{2}a_{0}^{(5-N)/2}$
in four different dimensions. Note that the horizontal axis starts
from the critical radius, beyond which the matter at the throat is
ordinary.}
\label{Fig01} 
\end{figure}

\begin{figure}[h]
\includegraphics[scale=0.3]{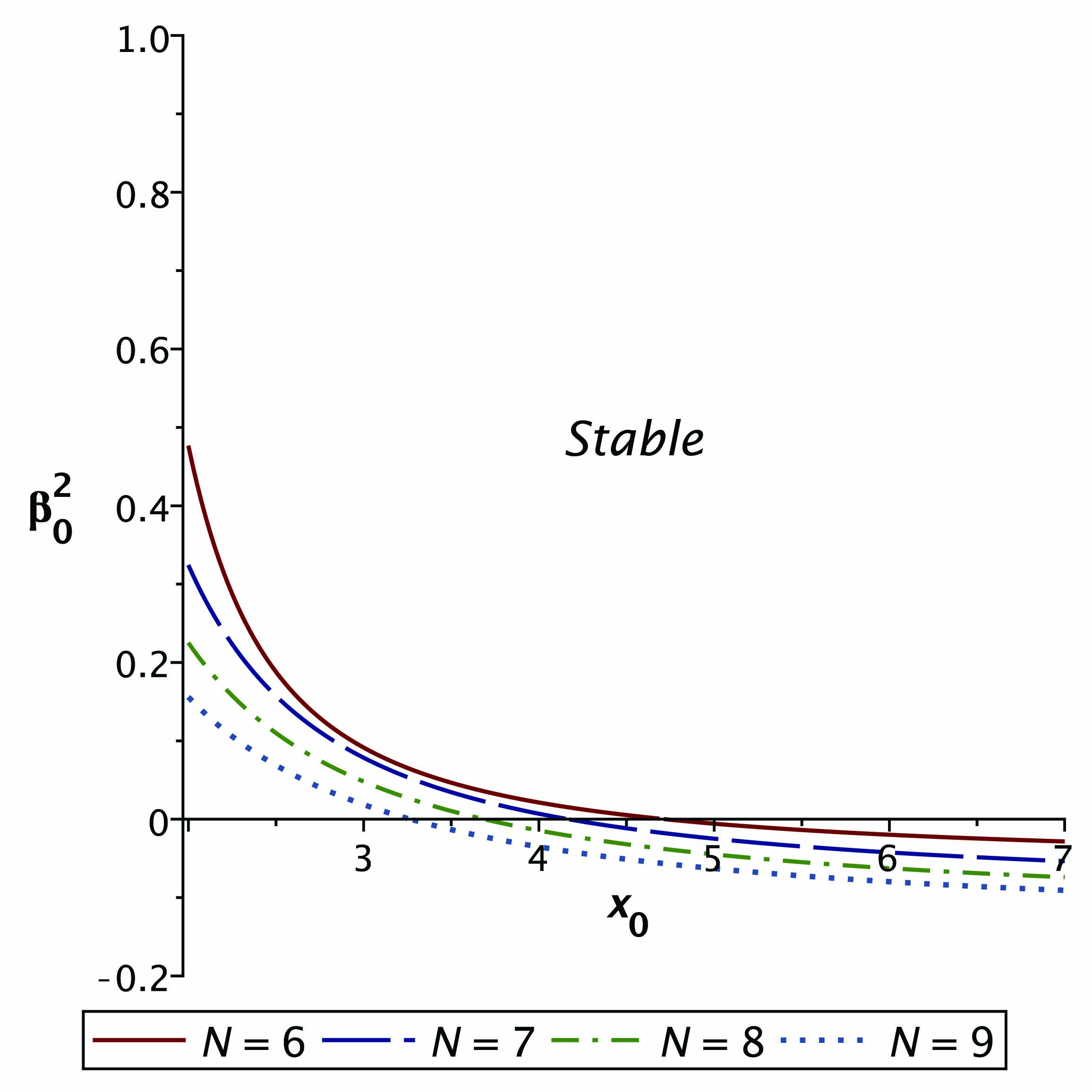}\caption{The stability diagrams for the non-black hole solution and a variable
equation of state where $\gamma_{0}=\left.-p_{0}^{\prime}\right|_{a=a_{D}}$
is chosen for the explicit radial dependency of the pressure. The
diagram plots $\beta_{0}^{2}$ against $x_{0}\equiv\omega^{2}a_{0}^{(5-N)/2}$
in four different dimensions. Note that the horizontal axis starts
from the critical radius, beyond which the matter at the throat is
ordinary.}
\label{Fig02} 
\end{figure}

After a radial perturbation is applied to the throat, its equation
of motion becomes 
\begin{equation}
\dot{a}^{2}+V\left(a\right)=v_{0}^{2},\label{equation of motion 2}
\end{equation}
in which $v_{0}$ is the initial velocity of the throat \cite{Zahra}.
For weak perturbation where $v_{0}^{2}\ll1,$ one may expand the potential
$V\left(a\right)$ near the equilibrium radius to write 
\begin{equation}
\dot{x}^{2}+V\left(a_{0}\right)+V^{\prime}\left(a_{0}\right)u+\frac{1}{2}V^{\prime\prime}\left(a_{0}\right)u^{2}+\mathcal{O}\left(u^{3}\right)\simeq v_{0}^{2},\label{expansion of the potential}
\end{equation}
in which $u=a-a_{0}.$ Explicit calculation shows that although $V\left(a_{0}\right)\neq0$,
yet $V^{\prime}\left(a_{0}\right)=0$, upon which (\ref{expansion of the potential})
becomes 
\begin{equation}
\dot{u}^{2}+\frac{1}{2}V^{\prime\prime}\left(a_{0}\right)u^{2}\simeq v_{0}^{2}-V\left(a_{0}\right).\label{40}
\end{equation}
Clearly, with $V^{\prime\prime}\left(a_{0}\right)>0$, $u$ will be
confined between the roots of $v_{0}^{2}=V\left(a_{0}\right)$, an
indication of the stability of the throat after the radial perturbation.
The potential $V\left(a\right)$ is of the form $V\left(a,\sigma\left(a\right)\right)$,
hence in finding $V^{\prime\prime}\left(a\right)$ one needs to know
$\sigma^{\prime}\left(a\right)$ and $\sigma^{\prime\prime}\left(a\right).$
In (\ref{energy conservation}), $\sigma^{\prime}\left(a\right)$
has been already found by using the energy conservation equation.
To calculate $\sigma^{\prime\prime}\left(a\right)$, we start from
$\sigma^{\prime}\left(a\right)$, and considering a variable equation
of state (EoS) \cite{Varela} for the matter at the throat ($p$ will
be a generic function of $\sigma$ and $a$ such that $p=p\left(\sigma,a\right)$),
we obtain 
\begin{equation}
\sigma^{\prime\prime}=\frac{N-2}{a^{2}}\left(\sigma+p\right)\left[N-1+\left(N-2\right)\left(\frac{\partial p}{\partial\sigma}\right)\right]-\frac{N-2}{a}\left(\frac{\partial p}{\partial a}\right).\label{sigma double-prime}
\end{equation}
At the equilibrium point, after the perturbation, one finds 
\begin{equation}
\sigma_{0}=-\frac{4\left(N-2\right)\tilde{c}_{2}}{3\kappa_{N}a_{0}^{3}}\sqrt{A_{0}+v_{0}^{2}}\left(3-A_{0}+2v_{0}^{2}\right),\label{sigma after perturbation}
\end{equation}
\begin{equation}
p_{0}=\frac{4\tilde{c}_{2}}{3\kappa_{N}a_{0}^{2}\sqrt{A_{0}+v_{0}^{2}}}\left[\frac{3}{2}A_{0}^{\prime}\left(1-A_{0}\right)+\frac{\left(N-5\right)}{a_{0}}\left(A_{0}+v_{0}^{2}\right)\left(3-A_{0}+2v_{0}^{2}\right)\right],\label{p after perturbation}
\end{equation}
\begin{equation}
\sigma_{0}^{\prime}=-\frac{N-2}{a_{0}}\left(\sigma_{0}+p_{0}\right),\label{sigma prime after perturbation}
\end{equation}
and 
\begin{equation}
\sigma_{0}^{\prime\prime}=\frac{N-2}{a_{0}^{2}}\left(\sigma_{0}+p_{0}\right)\left[N-1+\left(N-2\right)\beta_{0}^{2}\right]+\frac{N-2}{a}\gamma_{0},\label{sigma double-prime after perturbation}
\end{equation}
in which we have defined $\beta_{0}^{2}\equiv\left.\frac{\partial p}{\partial\sigma}\right\vert _{a=a_{0}}$
and $\gamma_{0}\equiv\left.-\frac{\partial p}{\partial a}\right\vert _{a=a_{0}}$.
The parameter $\beta_{0}$ is interpreted as the speed of sound within
the matter field at the throat. In case we encounter ordinary matter,
we shall have the condition $\beta_{0}^{2}$$\in\left(0,1\right)$
since the speed of light is taken as unity. For the non-black hole
solution, we have set $V^{\prime\prime}\left(a_{0}\right)$ equal
to zero and then plotted $\beta_{0}^{2}$ against the revearsed and
rescaled equilibrium radius $x_{0}\equiv\omega^{2}a_{0}^{(5-N)/2}$
for four different dimensions in Figs. \ref{Fig01} and \ref{Fig02},
where $\gamma_{0}=0$ and $\gamma_{0}=\left.-p_{0}^{\prime}\right|_{a=a_{D}}$,
respectively. Firstly, changing the static equilibrium radius from
$a_{0}$ to $x_{0}$ allows us to project the plots for different
dimensions onto a single diagram. The critical radius is now $x_{c}=2$
for every dimension, according to (\ref{critical radius}). However,
since $N>5$, the condition (\ref{critical radius}) now reads $x_{0}\geq x_{c}=2$.
Therefore, in Figs. \ref{Fig01} and \ref{Fig02}, the matter distributed
at the throat is ordinary post-$x_{c}$. Secondly, it is known that
the choice $\gamma_{0}=0$ evokes the well-known barotropic EoS, in
which the pressure is merely a generic function of the energy density,
i.e. $p=p\left(\sigma\right)$. Also, the choice $\gamma_{0}=\left.-p_{0}^{\prime}\right|_{a=a_{D}}$,
in which $a_{D}$ is the radius of discontinuity, is picked up deliberately
since it removes the discontinuity in the stability diagram of barotropic
TSWs \cite{Forghani}. Although the discontinuity radius in the case
we are studying here happens to locate behind the critical radius
at $x_{0}=2$, setting $\gamma_{0}=\left.-p_{0}^{\prime}\right|_{a=a_{D}}$
alters the behavior of the graph. In fact, we would like to see how
it affects the stability diagram of the TSW, for curosity. The discontinuity
radius is where $\sigma_{0}^{\prime}$ becomes null. The value of
the rescaled discontinuity radius is calculated as 
\begin{equation}
x_{D}=\frac{2\left(1+\sqrt{2N-1}\right)}{N-1}.\label{discontinuity radius}
\end{equation}
It is evident that the value of this radius for $N\geq6$ always happens
to be less the value of the critical radius at $2$.

In Figs. \ref{Fig01} and \ref{Fig02}, the regions of stability,
where $V^{\prime\prime}\left(a_{0}\right)$ is positive and the throat
is at a stable equilibrium, are marked. As it can be perceived from
Figs. \ref{Fig01} and \ref{Fig02}, for both barotropic fluid and
variable EoS fluid (and for the dimensions considered here), the TSW
could be radially stable in the physically meaningful range $\beta_{0}^{2}\in\left(0,1\right)$
beyond the critical radius $x_{c}=2$. Therefore, a TSW constructed
by a non-black hole vacuum solution in pure GB gravity, can maintain
ordinary matter and is stable against radial perturbations either
the fluid is supported by a barotropic or a variable EoS. Furthermore,
it is evident that for higher dimensions, it is more likely for the
TSW to be stable for both barotropic and variable EoSs. In addition,
for counterpart number of dimensions, a variable EoS TSW is more likely
to be stable than a barotropic TSW.

To complete our discussion in this part, let us argue the limit of
the real radius of TSW $a_{0}$. The critical radius in (\ref{critical radius})
becomes
\begin{equation}
a_{c}=\left(\frac{\mu}{4\tilde{c}_{2}}\right)^{\frac{1}{N-5}},\label{explicit critical radius}
\end{equation}
according to our definition $\omega^{2}\equiv\sqrt{\frac{\mu}{\tilde{c}_{2}}}$.
This critical radius, as the upper limit of the radius of the TSW,
is comparable to the radius of the event horizon of a Schwarzschild-Tangherlini
black hole in $N$ dimensions \cite{Tangherlini1}
\begin{equation}
r_{ST}=\mu^{\frac{1}{N-3}},\label{kappa}
\end{equation}
and is not in quantum scale (Note that the modified GB parameter $\tilde{c}_{2}$
has dimension $L^{2}$).

\begin{figure}[h]
\includegraphics[scale=0.3]{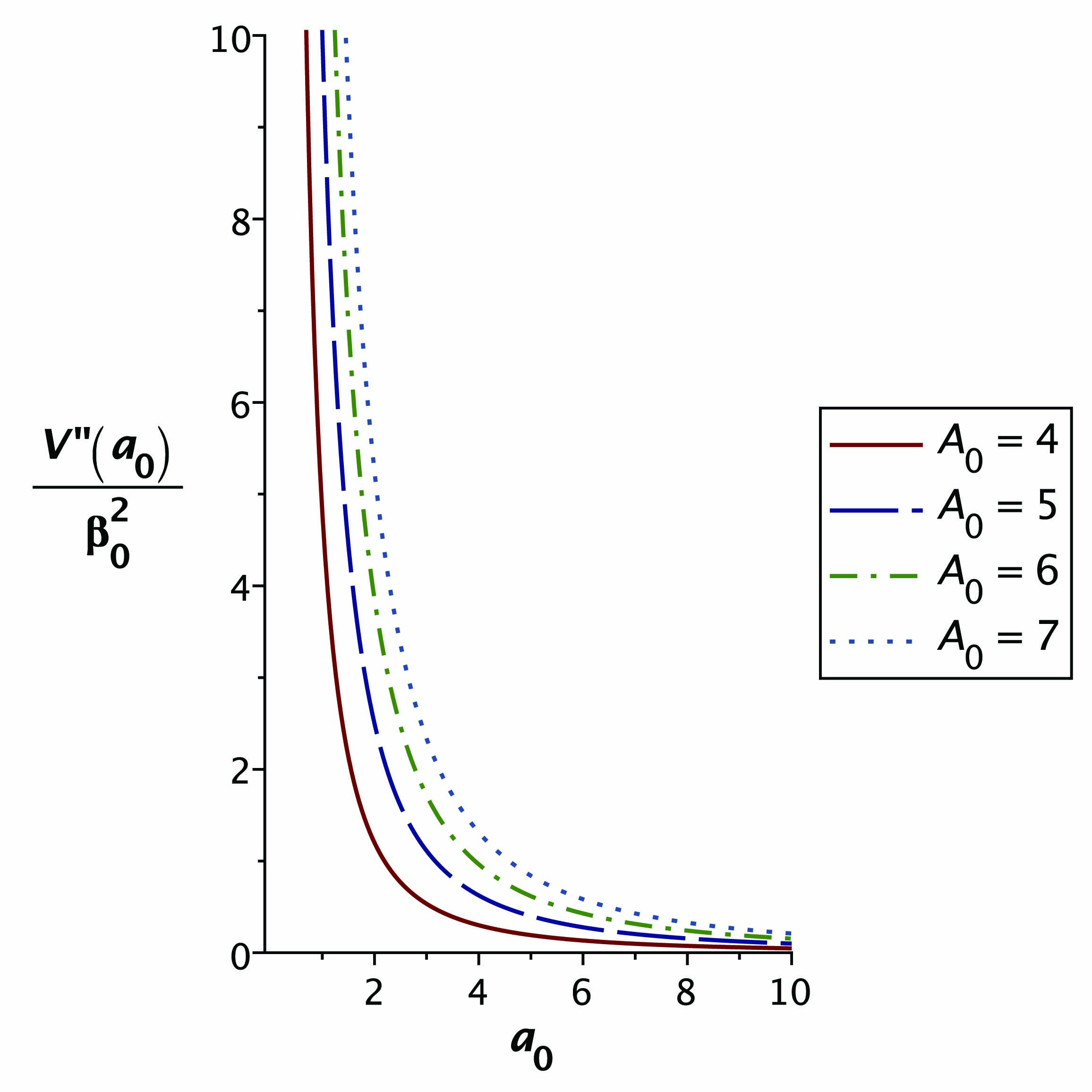} \caption{The plot indicates that $V^{\prime\prime}\left(a_{0}\right)/\beta_{0}^{2}$,
for all the values of $A_{0}$ considered here, is positive for any
radius of the throat.}
\label{Fig03} 
\end{figure}

For the cosmic string solution when $N=5$, our approach is slightly
different. Fig. \ref{Fig03} directly displays $V^{\prime\prime}\left(a_{0}\right)/\beta_{0}^{2}$
against $a_{0}$, for four different values of $A_{0}$. The values
of $A_{0}$ are chosen such that the TSW satisfies the WEC and the
matter is ordinary. As can be observed, for all the values of $A_{0}>3$,
$V^{\prime\prime}\left(a_{0}\right)/\beta_{0}^{2}$ is an absolutely
positive function of $a_{0}$, which, considering the physical condition
$\beta_{0}^{2}>0$, indicates that the TSW is stable against radial
perturbations. The figure suggests that this stability is stronger
for higher values of the constant $A_{0}$ and lower values of the
radius of the throat $a_{0}$.

\section{Conclusion \label{sec V}}

It has been decades that the cutting edge in wormhole studies has
been to find a wormhole-like structure that satisfies the known energy
conditions. It was known that in Einstein gravity, wormholes are supported
by exotic matter, which does not satisfy the energy conditions. TSWs,
which were introduced by Visser in 1989, opened doors to a wider class
of wormhole-like structures which also had this advantage that the
matter supported them was confined in a very limited area, say the
throat of the TSW. However, it was soon learned that the TSWs suffer
from the same exotic matter problem, as well. One way to bypass this
problem is to rely on modified theories of gravity, towards which
the Lovelock theory for its richness and simplicity is one of the
best choices. TSWs in third order Lovelock gravity have been studied
before and it was shown that under certain conditions they may satisfy
the energy conditions \cite{Dehghani}. In this study we challenged
the pure Lovelock gravity of order two, i.e. the pure GB gravity.
It was shown that for a non-black hole vacuum solution in $N>5$ dimensions,
if the throat's radius is less than a critical value (Eq. (\ref{critical radius})),
the thin-shell wormhole can be held together by ordinary matter. However,
for a black-hole solution in $N>5$ dimensions, the energy density
is always negative and hence the matter is always exotic. Also, it
was demonstrated that for the vacuum cosmic string solution in $5$
dimensions, the matter is ordinary under some certain conditions ($A_{0}\geq3$).
In continuation, we investigated the stability of such TSWs under
a radial perturbation by the standard linear stability analysis. It
was observed from Figs. \ref{Fig01} and \ref{Fig02} that for the
non-black hole TSW, there is a good possibility that the TSW is stable
and physical either the ruling EoS is barotropic or variable. However,
it is more likely for the TSW with a variable EoS to be stable than
a TSW with a barotropic EoS. Moreover, Fig. \ref{Fig03} suggests
that the cosmic string TSW is always stable when the matter is ordinary.

\section{Abbreviations}
\begin{lyxlist}{00.00.0000}
\item [{EoS:}] Equation of State
\item [{GB:}] Gauss--Bonnet
\item [{NEC:}] Null Energy Condition
\item [{TSW:}] Thin-Shell Wormhole
\item [{WEC:}] Weak Energy Condition
\end{lyxlist}

\section{Acknowledgment}

SDF would like to thank the Department of Physics at Eastern Mediterranean
University, especially the chairman of the department, Prof. İzzet
Sakallı for the extended facilities.

\end{document}